\def\bq{\begin{equation}}
\def\eq{\end{equation}}
\def\bqa{\begin{eqnarray}}
\def\eqa{\end{eqnarray}}
\def\bqb{\begin{eqnarray*}}
\def\eqb{\end{eqnarray*}}
\documentstyle[12pt]{article}\textwidth 16cm
\def\pr#1#2#3{ Phys. Rev. ${\bf{#1}}$ (#2) #3}

\def\pl#1#2#3{ Phys. Lett. ${\bf{#1}}$ (#2) #3 }

\def\np#1#2#3{ Nucl. Phys. ${\bf{#1}}$ (#2) #3}
\def\zp#1#2#3{ Z. Phys. ${\bf{#1}}$ (#2) #3}
\def\ijmp#1#2#3{ Int. J. Mod. Phys. ${\bf{#1}}$ (#2) #3}

\def\eg{{\it e.g.\/}}

\def\etal{{\it et.al.\/}}

\global\nulldelimiterspace = 0pt

\def\Bsl{\hbox{/\kern-.6700em$B$}} 
\def\Dsl{\hbox{/\kern-.6700em$D$}} 
\def\Wsl{\hbox{/\kern-.6700em$W$}} 

\def\roughly#1{\mathrel{\raise.3ex
    \hbox{$#1$\kern-.75em\lower1ex\hbox{$\sim$}}}}
\def\lsim{\roughly<}

\def\ol#1{\overline{#1}}

\def\L{ {\cal L }}
\def\O{ {\cal O }}

\def\mwd{M_W^2}

\def\lw{\lambda_W}

\def\mh2{m^2_H}

\begin{document}
\pagenumbering{arabic}
\thispagestyle{empty}
\hspace {-0.8cm} PM/95-31 \\
\hspace {-0.8cm} June 1995\\
\vspace {0.8cm}\\

\begin{center}
{\Large\bf Residual New Physics Effects in $e^+e^-$ and
$\gamma\gamma$ Collisions}
\footnote{Partially supported by the EC contract CHRX-CT94-0579.}
 \vspace{0.8cm}  \\
{\large  F.M. Renard}
\vspace {0.1cm}  \\
Physique
Math\'{e}matique et Th\'{e}orique,
CNRS-URA 768,\\
Universit\'{e} de Montpellier II,
 F-34095 Montpellier Cedex 5.\\

\vspace{1.5cm}

{\bf Contribution to 17th International Symposium on Lepton-Photon
Interactions, Beijing, Aug. 1995}\\

\vspace{0.5cm}

 {\bf Abstract}
\end{center}
\noindent
We present a set of studies concerning the description
of New Physics (NP) effects caracterized by a scale
much higher than the electroweak
scale. We show that both present experimental results and several types
of theretical considerations about the Standard Model (SM) and NP
suggest a hierarchy among the sectors in which NP manifestations should
appear. We concentrate on residual effects described by effective
lagrangians involving bosonic and/or heavy quark fields.
For each operator we propose an
unambiguous definition of the NP scale given by
the energy at which unitarity is saturated. We also consider the
possible existence of Higher Vector Bosons.
 We then study the tests realizable at
present and future colliders. We start from the analysis of the high
precision tests at Z peak, discussing separately the constraints
obtained from the light fermionic
sector and those due to the $b\bar b$ sector.
We then consider the process
$e^+e^-\to f\bar f$ at higher energies (LEP2 and NLC)
and we propose the "Z peak
subtracted representation" which allows to automatically
take into account Z peak constraints and to describe several types of NP
effects at any other energy. Applications to various NP effects (Higher
Vector Bosons, Technicolour resonances,
Anomalous Gauge Boson couplings) are
given. We then concentrate on several bosonic processes, $e^+e^-\to
W^+W^-$, $HZ$, $H\gamma$, and $\gamma\gamma$ collisions producing
boson pairs or a single Higgs. We study the sensitivities to the
various operators involved in the effective lagrangian and
we propose ways
to disentangle them. The NP scales which can be felt vary from a few
TeV at LEP2
up to 200 TeV in single H production at NLC.

\vspace{1cm}

\setcounter{page}{0}
\def\thefootnote{\arabic{footnote}}
\setcounter{footnote}{0}
\clearpage

\section{Introduction}

New Physics (NP) is a generic word designing unknown
dynamical features that
would solve all problems and deficiencies of the Standard Model (SM).
The nature of NP is an opened
question. Extensions of SM by a gauge group, extensions of the minimal
scalar Higgs sector, Supersymmetry, alternative descriptions of mass
generation, Grand Unified schemes and other new concepts
(superstrings),.... are popular examples.\par
Present experiments, in particular the high precision tests performed
at $Z$ peak and at low energies have excluded a modification of the SM
structure of the light leptons and quarks at a few \underline{per
mille} level\cite{LEP1}.
This can be interpreted by saying that NP is caracterized
by a scale $\Lambda_{NP}$ much larger than $M_Z$, at least in this light
fermionic sector. On another hand the most acute theoretical
problem of SM concerns the mass generation mechanism\cite{Haber},
i.e. the scalar
sector which generates the longitudinal $W_L$, $Z_L$ states as well
as the heavy ($t$, $b$)
quark sector. These facts are driving us to
the search for NP manifestations in these latter sectors. \par
We shall stick to the assumption that $\Lambda_{NP}$ is higher than the
energies reachable by present and nearby future experiments, so that no
new particles can be produced. The only NP manifestations should then
consist in modifications of the interactions among usual particles that
are called residual effects. The net result is the appearence
of  modified SM couplings among usual particles, of departures
from the SM values of the standard gauge couplings, and the
existence of new
coupling forms.
They are described by an effective
lagrangian $\L_{eff}$ corresponding to the integration of all heavy (of
order $\Lambda_{NP}$) degrees of freedom. The form, a priori unknown,
of $\L_{eff}$ is restricted by general principles like Lorentz
and $U(1)_{EM}$ invariance and by the dimension of the operators used
to construct it. In practice the high value of the NP scale should
favor the $dim=6$ operators\cite{Buchmuller}.
Each operator is associated to a coupling
constant and we have shown \cite{unit}, \cite{uni2},\cite{model},
that its value is related to the energy at
which unitarity is saturated. This energy value can be considered as an
unambiguous definition of the NP scale as it should correspond to the
threshold for new particle creation.\par
Further assumptions can reduce the number
of operators.  Recently it has been advocated \cite{DeR}
that in order to preserve
all "good features" of the SM, a broken gauge invariance scenario has
to be kept. Before Symmetry Breaking (SB) applies, $L_{eff}$ should
be $SU(2)\times U(1)$ gauge invariant. When the dimension is
limited as mentioned above, this is an important restriction, otherwise
it is always possible to write any Lorentz and $U(1)_{EM}$ invariant
lagrangian into an $SU(2)\times U(1)$ gauge invariant one by taking
suitable combinations of scalar fields with other fields\cite{g.i.}.
At this point let us
however notice that there exist possibilities of
extended gauge invariant pictures which do not fall into this
class of structures. At low energies they induce explicit SB
breaking. See for example the so-called
strong $SU(2)_V$ extensions\cite{SU2V}.
For simplicity we shall not consider these
possibilities in this report.\par
We have also studied \cite{dyn},\cite{model},
how $\L_{eff}$ is generated by specific
NP structures, for example
when one integrates out the effects of new heavy
fermions or new heavy bosons. We have observed that it seems easier to
generate stronger anomalous couplings for Higgs bosons than for gauge
bosons. A hierarchy is then appearing for NP effects which puts ahead
the scalar sector, followed by the gauge boson sector and the heavy
quark sector and finally the light quark sector which is already very
strongly constrained by present experiments.\par
Another possible manifestation of NP in the bosonic sector is through
the existence of higher vector bosons. The effects of a higher $Z'$ are
of two kinds, direct modifications of the $Z$ couplings through $Z-Z'$
mixing and addition of $Z'$ exchange diagrams,
that we shall also describe
in an effective way as a modification of the SM amplitudes for photon
and Z exchange. We have recently reexamined both of these effects
\cite{ZZp}, \cite{ZZplep2}, \cite{RVtraj}.\par
After shortly presenting, in Section 2,
the tools that we used for
treating these various NP manifestations, I discuss  in Section 3
the tests that can
be done at present and future colliders with
the expected sensitivities to
each type of effect and I present, in Section 4,
the landscape of NP which
comes as an ouput.\par

\section{Higher Vector Bosons and Effective Lagrangians}

{\bf a) Higher  Neutral Vector Bosons $Z'$}

The existence of Higher Vector Bosons is
predicted in various extensions of the SM ($E_6$ grand unified
schemes, Right-Left symmetry) or
alternatives (compositeness inspired schemes).
No direct production has yet occured (a limit of the order of
$M_V > 500 GeV$ has been set at Fermilab\cite{Zpfnal}).\par
We are interested in the indirect limits that can be set on neutral
vector bosons (generically called $Z'$) in
$e^+e^-$ collisions. They come in two ways. Firstly, a $Z'$
exchange diagram can be added to the $\gamma$ and $Z$ one in $e^+e^-$
annihilation into fermion pairs. Its effect will directly depend on the
vector and axial couplings of the $Z'$ and on its mass $M_{Z'}$.
We have shown \cite{RVtraj} that this effect can be
described as a
modification of the initial and final $\gamma, Z$ couplings and treated
together with the 1-loop description of the SM amplitude at any
$q^2$. I shall discuss the results in Section 3.\par
 However the $Z'$ may be only weakly coupled  to
fermion-antifermion. This arises in several alternative models for mass
generation (Technicolour, Compositeness, $SU(2)_V$ extensions,...). In
this case their existence can be searched through their bosonic
couplings ($W^+W^-$) or through several types of modifications they may
induce on $\gamma, W, Z$ properties. For example
$Z-Z'$ mixing through loop effects
or through genuine mass-mixing can modify Z couplings.
As they
have been
very accurately measured at LEP1, strong constraints on the mixing
angle $\theta_M$ has been established \cite{ZZp}.
We shall come back to these quantitative limits in Section 3.\par
\vspace{0.5cm}
\noindent
{\bf b) Effective Lagrangian in the Bosonic Sector}

Following the assumption that the strongest NP residual effects lie in
the bosonic sector, and that they are caracterized by a high scale, we
have concentrated our study on operators of $dim=6$, $SU(2)\times U(1)$
gauge invariant and involving only $W, Z, \gamma$ and $H$. The full
list involves 17 operators\cite{Hag}, \cite{model}.
They have been first classified according to the
way they affect the various sectors. Eight non blind operators, 4
CP-conserving ones $\O_{DW}$, $\O_{DB}$, $\O_{BW}$, $\O_{\Phi1}$
and 4 CP-violating ones $\ol{\O}_{W}$, $\ol{\O}_{W\Phi}$,
$\ol{\O}_{B\Phi}$,
$\ol{\O}_{BW}$
affect the light fermionic sector and are severely
constrained by Z peak as well as low energy measurements.
Two superblind ones $\O_{\Phi2}$, $\O_{\Phi3}$,
do not
affect fermions neither pure gauge bosons couplings
but only Higgs couplings to themselves or to gauge bosons, and
are therefore difficult to observe experimentally. We shall
concentrate on the seven
blind ones which are not affecting the light fermions but produce
anomalous gauge boson couplings and anomalous Higgs-gauge boson
couplings. We shall also consider $\O_{\Phi2}$ which can affect the
process $e^+e^-\to HZ$. This set of eight operators is listed below.

The first one produce only anomalous gauge
boson couplings
\bqa
\O_W &= & {1\over3!}\left( \overrightarrow{W}^{\ \ \nu}_\mu\times
  \overrightarrow{W}^{\ \ \lambda}_\nu \right) \cdot
  \overrightarrow{W}^{\ \ \mu}_\lambda
\eqa

The next two ones produce both anomalous gauge
boson couplings and anomalous Higgs-gauge boson
couplings
\bqa
\O_{W\Phi} & = & i\, (D_\mu \Phi)^\dagger \overrightarrow \tau
\cdot \overrightarrow W^{\mu \nu} (D_\nu \Phi) \ \ \  , \ \
 \\[0.1cm]
\O_{B\Phi} & = & i\, (D_\mu \Phi)^\dagger B^{\mu \nu} (D_\nu
\Phi)\ \ \  , \
\eqa

\noindent
whereas the four other ones produce anomalous Higgs couplings but no
pure gauge-boson couplings.
\bqa
\O_{UW} & = & \frac{1}{v^2}\, (\Phi^\dagger \Phi - \frac{v^2}{2})
\, \overrightarrow W^{\mu\nu} \cdot \overrightarrow W_{\mu\nu} \ \
\  ,  \ \ \\[0.1cm]
\O_{UB} & = & \frac{4}{v^2}~ (\Phi^\dagger \Phi -\frac{v^2}{2})
B^{\mu\nu} \
B_{\mu\nu} \ \ \  , \ \ \\[0.1cm]
 \ol{\O}_{UW} & = & \frac{1}{v^2}\, (\Phi^\dagger
\Phi)
\, \overrightarrow W^{\mu\nu} \cdot
\widetilde{\overrightarrow W}_{\mu\nu} \ \
\  ,  \ \ \\[0.1cm]
\ol{\O}_{UB} & = & \frac{4}{v^2}~ (\Phi^\dagger \Phi )
B^{\mu\nu} \
\widetilde{B}_{\mu\nu} \ \ \  \ \ \
\eqa
The last two ones are CP-violating.
We shall also consider the superblind operator
\bq
\O_{\Phi2} ~=~4
\partial_\mu(\Phi^\dagger \Phi)\partial^\mu(\Phi^\dagger \Phi)
\ \ \ \ , \
\eq
Its effect is to induce
a wave function renormalization to the Higgs field.
 Note also that $\O_{\Phi2}$, $\O_W$, $\O_{UW}$
and $\ol{\O}_{UW}$ are custodial $SU(2)_c$ invariant.\par
The effective lagrangian will be written
\bqa
\L_{NP} & = & \lw \frac{g}{\mwd}\O_W +
{f_B g\prime \over {2M^2_W}}\O_{B\Phi} +{f_W g \over
{2M^2_W}}\O_{W\Phi}\ \  + \nonumber \\
\null & \null &
d\ \O_{UW} + {d_B\over4}\ \O_{UB} +\ol{d}\ \ol{\O}_{UW} +
{\ol{d}_B\over4}\ \ol{O}_{UB} +\frac{f_{\Phi2}}{v^2} \O_{\Phi 2}\ \ \ \
\ \ ,  \
\eqa
This NP lagrangian is often written elsewhere as

\bq   L = \Sigma_i {\bar f_i \over \Lambda^2_{NP}}O_i  \eq

$\Lambda_{NP}$ being the NP scale.
However in absence of a well defined NP
dynamics, the normalization of $\Lambda_{NP}$ is not well defined,
and in
phenomenological analyses it has to be fixed to a certain value (often
taken as $1TeV$) before discussing the magnitude of the $\bar f_i$.\par
We have defined\cite{unit} a phenomenological
scale which is free of this
ambiguity. For $dim>4$ operators the amplitudes grow with the energy so
that we define a scale $\Lambda_{th}(f_i)$
as the value of the energy at which the
unitarity limit is reached for a given operator, the lagrangian
being now written
as

\bq   L = \Sigma_i  f_i O_i  \ \  .\ \ \eq

This scale $\Lambda_{th}$ has precisely the physical
meaning of the NP scale $\Lambda_{NP}$ as it
should precisely be the energy value
corresponding to the threshold for new particle production (which cures
the unitarity saturation). So in the following we shall identify these
two scales and for each operator we shall refer to
$\Lambda_{NP} \equiv \Lambda_{th}(f_i)$. It is also
the energy at which $\L_{eff}$ ceases
to be meaningful. So for each operator $\O_i$, each value of $f_i$ is
associated to a $\Lambda_{NP}$ through a well-defined relation. An
example extracted from the results obtained
in \cite{unit} for the operator
$\O_W$ with the normalization in eqs.(1),(9) is

\bq
\Lambda_{th}(\lambda_W) = \sqrt{19{M^2_W\over|\lambda_W|}}  \eq

It shows in particular that the sensitivity to low
values of $f_i$ means a sensitivity to high scales $\Lambda_{NP}$.\\
\vspace{0.3cm}

\noindent
{\bf c) Effective Lagrangian for the Heavy Quark Sector}

Contrarily to the case of the light fermionic sector, the heavy quark
sector $(t, b)$ has not yet been tested with high accuracy. At LEP1, the
recent measurements of the $Zb\bar b$ width show room for non standard
effects at the few percent level (as opposed to the per mille level in
the light sector)\cite{LEP1}.
The heavy top ($m_t\simeq 2M_W$) may precisely open
a window on the mass generation mechanism. Anomalous top properties
reflect in the $Zb\bar b$ coupling through the
1-loop correction to this vertex. In the SM case a
large effect proportional to $(m_t/M_Z)^2$ is already appearing. NP
effects involving the top can also be similarly enhanced. If the mass
generation mechanism is at the origin of these effects, we expect
that the corresponding effective lagrangian involves at least one $t_R$
field. We have done\cite{topbb} a classification of all such $dim=6$,
$SU(2)\times U(1)$ gauge invariant operators and we found 28 ones, 14 of
them involving $t_R$ fields. We concentrated on these 14 ones that
consist in 7 two-quark operators and 7 four-quark ones.\par

 \underline {Four-quark operators}
\bqa
\O_{qt} & = & (\bar q_L t_R)(\bar t_R q_L) \ \ \ , \ \\[0.1cm]
\O^{(8)}_{qt} & = & (\bar q_L \overrightarrow\lambda t_R)
(\bar t_R \overrightarrow\lambda q_L) \ \ \ ,\ \\[0.1cm]
\O_{tt} & = & {1\over2}\, (\bar t_R\gamma_{\mu} t_R)
(\bar t_R\gamma^{\mu} t_R) \ \ \ , \ \\[0.1cm]
\O_{tb} & = & (\bar t_R \gamma_{\mu} t_R)
(\bar b_R\gamma^{\mu} b_R) \ \ \ , \ \\[0.1cm]
\O^{(8)}_{tb} & = & (\bar t_R\gamma_{\mu}\overrightarrow\lambda t_R)
(\bar b_R\gamma^{\mu} \overrightarrow\lambda b_R) \ \ \ , \
\eqa
\bqa
\O_{qq} & = & (\bar t_R t_L)(\bar b_R b_L) +(\bar t_L t_R)(\bar
b_L b_R)\ \ \nonumber\\
\null & \null & - (\bar t_R b_L)(\bar b_R t_L) - (\bar b_L t_R)(\bar
t_L b_R) \ \ \ , \ \\[0.1cm]
\O^{(8)}_{qq} & = & (\bar t_R \overrightarrow\lambda t_L)
(\bar b_R\overrightarrow\lambda b_L)
+(\bar t_L \overrightarrow\lambda t_R)(\bar b_L
\overrightarrow\lambda  b_R)
\ \nonumber\\
\null & \null &
- (\bar t_R \overrightarrow\lambda b_L)
(\bar b_R \overrightarrow\lambda t_L)
- (\bar b_L \overrightarrow\lambda t_R)(\bar t_L
\overrightarrow\lambda   b_R)
\ \ \  . \
\eqa\\
 \underline {Two-quark operators}
\bqa
\O_{t1} & = & (\Phi^{\dagger} \Phi)(\bar q_L t_R\widetilde\Phi +\bar t_R
\widetilde \Phi^{\dagger} q_L) \ \ \ ,\ \\[0.1cm]
\O_{t2} & = & i\,\left [ \Phi^{\dagger} (D_{\mu} \Phi)- (D_{\mu}
\Phi^{\dagger})  \Phi \right ]
(\bar t_R\gamma^{\mu} t_R) \ \ \ ,\\[0.1cm]
\O_{t3} & = & i\,( \widetilde \Phi^{\dagger} D_{\mu} \Phi)
(\bar t_R\gamma^{\mu} b_R)-i\, (D_{\mu} \Phi^{\dagger}  \widetilde\Phi)
(\bar b_R\gamma^{\mu} t_R) \ \ \ ,\\[0.1cm]
\O_{D t} &= & (\bar q_L D_{\mu} t_R)D^{\mu} \widetilde \Phi +
D^{\mu}\widetilde \Phi^{\dagger}
(\ol{D_{\mu}t_R}~ q_L) \ \ \ , \\[0.1cm]
\O_{tW\Phi} & = & (\bar q_L \sigma^{\mu\nu}\overrightarrow \tau
t_R) \widetilde \Phi \cdot
\overrightarrow W_{\mu\nu} + \widetilde \Phi^{\dagger}
(\bar t_R \sigma^{\mu\nu}
\overrightarrow \tau q_L) \cdot \overrightarrow W_{\mu\nu}\ \ \
,\\[0.1cm]
\O_{tB\Phi}& = &(\bar q_L \sigma^{\mu\nu} t_R)\widetilde \Phi
B_{\mu\nu} +\widetilde \Phi^{\dagger}(\bar t_R \sigma^{\mu\nu}
 q_L) B_{\mu\nu} \ \ \ ,\\[0.1cm]
\O_{tG\Phi} & = & \left [ (\bar q_L \sigma^{\mu\nu} \lambda^a t_R)
\widetilde \Phi
 + \widetilde \Phi^{\dagger}(\bar t_R \sigma^{\mu\nu}
\lambda^a q_L)\right ] G_{\mu\nu}^a  \ \ \ .
\eqa
We temporarily (the obtention of unitarity constraints
is in progress \cite{dirtop}) define the lagrangian in this sector as
\bq
\L =  \sum_i { \bar f_i \over \Lambda^2_{NP}}\,\O_i \ \ \ , \
\eq
$\Lambda_{NP}$ being the NP scale and $\bar f_i$ the dimensionless
coupling of the ~operator $\O_i$.
The effects in $Z$ peak physics has been studied in \cite{topbb} and
reported in Section 3. The analysis of the effects in direct top
production
is in progress\cite{dirtop}.

\section{Tests at present and future colliders}

We now successively report on the results obtained with the above
tools for the present
LEP/SLC range and beyond (LEP2 and NLC).\\
\vspace{0.3cm}

{\bf 3a. The Light l,q Sector at LEP1/SLC}

It is important to discuss how far the high precision tests done at Z
peak with fermionic processes can be used to test the bosonic sector.
In order to achieve this goal it is essential to use a description of
the Z exchange processes which is sufficiently general in order to
account for possible NP effects but also in order
to cover in an accurate way the
SM radiative correction effects.
For this reason  the usual description \cite{AB} of the effective
$Z$ exchange amplitude in $e^+e^- \to f \bar f$ has been
somewhat generalized \cite{vertex}.\par

The charged leptonic processes $e^+e^- \to Z \to l^+l^-$ is
described by the usual two parameters $\epsilon^l_1 , \bar s^2_l$
that can be  experimentally measured through two "good" observables,
the leptonic $Z$ width $\Gamma_l$, and  $A_l$ taken as
the polarized asymmetry $A_{LR}$ or
the $\tau$ asymmetry or through the unpolarized
forward-backward asymmetry $A_{FB,l}={3\over4}A^2_l$.\par
The light quark processes $e^+e^- \to Z \to q \bar q$ are described
assuming universality for the first two families, i.e. $u=c$
and $d=s$. In this case
we have defined 4 parameters $\epsilon^{u,d}_1$, $\bar s^2_{u,d}$
(actually in \cite{vertex} we have frequently used their combinations
called  $\delta^{(1)}_{u,d}$ and $\delta'_{u,d}$).
They could in principle be determined by the four observables
 $\Gamma_{u,d}$ or $\Gamma_{c,s}$ and $A_{u,d}$ or $A_{c,s}$.
In practice the situation is slightly less simple as one can measure in
an accurate way only $\Gamma_4$ or the combination $D$ defined
in \cite{D}
and at a weaker level maybe also $\Gamma_c$.\par
 Asymmetry factors $A_q$ are involved in the forward-backward
asymmetries $A_{FB,q}={3\over4}A_lA_q$ but can only be measured
with a sufficient accuracy through polarized
$e^{\pm}$ beams\cite{LEPpol}
with $A^{pol(q)}_{FB}= {3\over4}A_q$ for q=c and at a weaker accuracy
for q=s.\par

Constraints on Higher Vector Bosons and on anomalous interactions in
the bosonic sector have been established on the basis of these $Z$ peak
measurements. \par
In the case of \underline{$Z-Z'$ mixing effects},
model-independent constraints
have been obtained \cite{ZZp}, \cite{ZZplep2}
on the product of the mixing angle $\theta_M$ by
the fermionic $Z'$ couplings. For specific models (like $E_6$ or $LR$
symmetry, or alternative models) in which these $Z'$ couplings are
fixed,
upper limits have been found for $\theta_M$ that lie at the few per
mille level, see Table 1a.
\begin{center}
\begin{tabular}{|c|c|c|c|} \hline
\multicolumn{4}{|c|}{Table 1a: Upper bounds on the mixing angle
$\theta_M$.} \\ \hline
\multicolumn{1}{|c|}{$\Psi$} &
  \multicolumn{1}{|c|}{$\eta$} &
   \multicolumn{1}{|c|}{$\chi$} &
    \multicolumn{1}{|c|}{$LR$}   \\ \hline
 -0.007\ \ +0.002 & -0.012\ \ +0.005
& -0.010\ \ +0.002 & -0.005\ \ +0.002 \\ \hline
\end{tabular}
\end{center}
\noindent
 In some "constrained" cases this mixing angle is related
to the ratio $M^2_Z/M^2_{Z'}$ and in Table 1b
the limits have been translated into
lower limits for $M_{Z'}$ \cite{ZZp}, \cite{ZZplep2}.\par
\begin{center}
\begin{tabular}{|c|c|c|c|c|c|} \hline
\multicolumn{6}{|c|}{Table 1b: Lower bounds for Z' masses in TeV} \\ \hline
\multicolumn{1}{|c|}{$\eta$} &
  \multicolumn{1}{|c|}{$\chi$} &
   \multicolumn{1}{|c|}{$LR$} &
    \multicolumn{1}{|c|}{$Y$ } &
     \multicolumn{1}{|c|}{$Y_L(\lambda^2_Y=1/4)$} &
      \multicolumn{1}{|c|}{$Z^*$ } \\ \hline
 0.6 & 0.6 & 1.1 & 1.1 & 1.7 & 1.1 \\ \hline
\end{tabular}
\end{center}
\noindent
Concerning the \underline{effective bosonic lagrangian},
I will just recall that
the four non blind operators $\O_{DW}$, $\O_{DB}$, $\O_{BW}$,
$\O_{\Phi 1}$  contribute directly (at tree level) to the
$\epsilon_i$ parameters. Consequently the constraints on the coupling
constants are very strong\cite{DeR}, \cite{Hag}, i.e.
\bq  |\bar f_i{M^2_Z\over\Lambda^2_{NP}}|
\lsim O(10^{-2}\ \ to\ \ 10^{-3}) \eq
when one treats these operators one by one.\par

Blind operators affect the LEP1 parameters only at 1-loop. This raises
the usual technical problems of loop computations with effective
(non renormalizable) lagrangians\cite{Burgess}.
About this point we have just noticed
the fact that the domain of
integration
corresponding to the divergent part may correspond to
a strong coupling regime (and even overpass
the unitarity limit) so that non-perturbative effects should in
principle be taken into account. This weakens the power of the
constraints that has been derived from perturbative analyses.
Nevertheless they give an orientation. In any case, because of the
loop factor ${\alpha \over {4\pi}}$ the constraints are
much weaker than in the case of nonblind operators
\bq  |\bar f_i{M^2_Z\over\Lambda^2_{NP}}| \lsim O(1\ \ to
\ \ 10^{-1})\eq
for example\cite{Hag}\par
\bq  |\lambda_W| \lsim 0.6 \ \ \ . \ \  \eq
 Through unitarity relations \cite{unit} this limit corresponds
to a rather low NP scale of 0.35  TeV and this illustrates
the point just mentioned above about the limit of validity of the
1-loop constraints.\par
We have also computed\cite{topbb} the leading contributions at 1-loop
of the anomalous top
interactions described in Section 2 to the $\epsilon_i$ parameters.
In fact only 4 operators
$\O_{t2}$, $\O_{Dt}$, $\O_{tW\Phi}$, $\O_{tB\Phi}$
in this
list contribute here and the result, for $\Lambda_{NP} = 1 TeV$, is

\bq  -0.3 \lsim \bar f_{t2} \lsim +0.3
\ \ \ \ \ \ -1.1 \lsim \bar f_{Dt} \lsim +1.1 \ \ \ ,\eq
\bq  -0.27 \lsim \bar f_{tW\Phi} \lsim +0.47
\ \ \ \ \ \ -0.27 \lsim \bar f_{tB\Phi} \lsim +0.43 \ \ \ , \eq
which constitutes already very stringent constraints on this sector.

\vspace{0.3cm}

{\bf 3b. The Heavy Quark Sector at LEP1/SLC}

In this sector the only process available at
Z peak is $e^+e^- \to Z \to b \bar b$.
It is described by two additional
parameters \cite{Comelli}
that we identify through the departures from universality
with the two first families:
\bq   \delta g_{Vb} = \delta g_{Vd} + \delta g^{Heavy}_{Vb} \eq
\bq   \delta g_{Ab} = \delta g_{Ad} + \delta g^{Heavy}_{Ab} \eq
that can be determined through the two new observables
\bq \Gamma_b = \Gamma_d[1+\delta_{bV}]
\ \ \ \ \ \ \ \ A_b = A_d[1+\eta_{b}] \eq
 measurable through
\bq   R_b = {\Gamma_b\over \Gamma_{had}}
 \ \ \ \ \ \ \ A^{pol(b)}_{FB} = {3\over4}A_b \eq

The parameters $\delta_{bV}$ and $\eta_{b}$ are
essentially representing the left-handed and the right-handed types of
NP corrections \cite{Comelli}. These corrections can appear from various
sources (anomalous heavy quark couplings, anomalous gauge boson
couplings, anomalous Higgs couplings, new particle exchanged
like charged Higgses and supersymmetric partners,.....).
We have shown that various models lie on
different trajectories in the $\delta_{bV}, \eta_b$ plane. See Fig.
2-6 of \cite{Comelli}.\par
Recently we have computed the leading contributions to the $Zb\bar b$
vertex of the operators describing the anomalous gauge boson and
top interactions
listed in Section 2. In the case of anomalous gauge boson interactions
only 2 operators, namely $\O_{W\Phi}$ and $\O_{B\Phi}$, contribute at
this level \cite{RV}. So this process provides an interesting way of
disentangling them from the whole set of blind operators.\par
 In the case of anomalous top interactions, among
the 14 operators only seven of them contribute and the
results are summarized in Table 2, where the blanks indicate no
contribution from the corresponding operator.
\noindent
\begin{center}
\begin{tabular}{|c|c|c|c|c|} \hline
\multicolumn{5}{|c|}{Table 2: Contributions of "top" operators
to $Z$ peak physics.}\\[.1cm] \hline
\multicolumn{1}{|c|}{Operator} &
  \multicolumn{1}{|c|}{$\epsilon^{(NP)}_1$} &
   \multicolumn{1}{|c|}{$\epsilon^{(NP)}_3$ } &
     \multicolumn{1}{|c|}{$\delta^{(NP)}_{bv}$} &
       \multicolumn{1}{|c|}{$\eta^{(NP)}_b /\delta^{(NP)}_{bv}$}
          \\[0.1cm] \hline
  $\O_{qt}$ & \null & \null & $-2.1\times 10^{-3}
\bar f_{qt}$ & $0.068$ \\[0.1cm]
  $\O^{(8)}_{qt}$ & \null & \null & $-1.1\times
         10^{-2}\bar f^{(8)}_{qt}$ & 0.068 \\[0.1cm]
  $\O_{t2}$ & $-1.1 \times 10^{-2}\bar f_{t2} $ & \null & $2.1\times
       10^{-3}\bar f_{t2}$  & $0.068$ \\[0.1cm]
  $\O_{Dt}$ &$-2.8 \times 10^{-3} \bar f_{Dt}$ & \null &$-4.8\times
      10^{-3}\bar f_{Dt}$   & $0.068$ \\[0.1cm]
  $\O_{qq}$ & \null & \null & $1.7\times 10^{-5}
\bar f_{qq}$  & $0.03$ \\[0.1cm]
  $\O^{(8)}_{qq}$ & \null & \null & $9.1\times
         10^{-5}\bar f^{(8)}_{qq}$ & $0.03$ \\[0.1cm]
$\O_{tb}$ & \null & \null & $-2.3 \times 10^{-3}\bar f_{tb}$ &
$-2.068$ \\ [0.1cm]
$\O_{tW\Phi}$ & \null & $-6.0\times 10^{-3}\bar f_{tW\Phi}$ & \null &
\null \\[0.1cm]
  $\O_{tB\Phi}$ & \null & $-6.6\times 10^{-3}\bar f_{tB\Phi}$ & \null &
  \null \\ \hline
 \end{tabular}
\end{center}
\noindent

We remark that
the operators
$\O_{qt}$, $\O^{(8)}_{qt}$, $\O_{t2}$ and  $\O_{Dt}$ give purely
left-handed contributions, whereas
on the contrary, the operator
$\O_{tb}$ generates  a pure right-handed one.
Finally, $\O_{qq}$ and
$\O^{(8)}_{qq}$  generate only anomalous magnetic
moment-type couplings for both, $Zb\bar b$ and $\gamma b\bar b$
couplings.

The results presently available on $\Gamma_b$ alone from
LEP
and SLC \cite{LEP1}, leading to
\bq
\delta^{(NP)}_{bV}=(+1.93\pm1.08)\times10^{-2} \ \ \ .
\eq
give the constraints for $\Lambda_{NP}=1TeV$
\bq  -15 \lsim \bar f_{qt} \lsim  -4 \ \ \
\ \ \  -3 \lsim \bar f^{(8)}_{qt} \lsim  -0.7  \ \ \
\ \ \ -6 \lsim \bar f_{Dt} \lsim  -2  \ \ \
\ \ \ +4 \lsim \bar f_{t2} \lsim  +15  \ \ \ ,\eq
\bq  -14 \lsim \bar f_{tb} \lsim  -4  \ \ \
\ \ \ 0.5\times10^{+3} \lsim \bar f_{qq} \lsim 2\times10^{+3} \ \ \
\ \ \ 10^{+2} \lsim \bar f^{(8)}_{qq} \lsim 4\times10^{+2} \ .
\eq \par

The very loose limit on $\bar f_{qq}$ and $\bar f^{(8)}_{qq}$ is due
to the
presence of an $m_b/m_t$ factor for the effect of this type of
magnetic coupling called
$\delta \kappa^Z$ in \cite{topbb} .
It corresponds to a $\delta \kappa^Z$
value of the order of 0.1. We have looked whether it could be possible
to measure separately the magnetic $\gamma b \bar b$ and $Zb \bar b$
couplings by performing measurements outside
the Z peak. An analysis\cite{topbb}
of the process $e^+e^-\to b\bar b$ going
through photon and Z exchange showed us that
an accuracy of one percent below the Z
peak would allow the determination of
$\delta \kappa^\gamma$ at the same level of 0.1 .

The most interesting  result in Table 2 is given by its last column
which indicates that the ratio $\eta^{(NP)}_b
/\delta^{(NP)}_{bv}$ provides a very strong signature
for discriminating between the left-handed, right-handed and
the anomalous magnetic $Zb\bar b$ vertex. Note that if a single
operator dominates,  the ratio $\eta^{(NP)}_b
/\delta^{(NP)}_{bv}$ is independent of the magnitude of its
coupling and depends
only on the nature of the induced $Zb\bar b$ vertex.\par

It should be stressed that the large and negative
$\eta^{(NP)}_b /\delta^{(NP)}_{bv}$ ratio would be a rather peculiar
signature of the $\O_{tb}$ operator.
This result would be orthogonal to the  expectations for the minimal
supersymmetric SM. Here, in fact, the trend would be that of
\underline{positive} $\eta^{(NP)}_b$ (of order one percent) for
positive $\delta^{(NP)}_{bv}$. However, this prediction would be
necessarily acompanied by the discovery of suitably light
supersymmetric particles, like \eg\@ a light chargino and/or a
light neutral Higgs.

\section{$e^+e^-\to f \bar f$ at LEP2 and NLC}

We were then interested in the following question. Having as an input
the result of the high precision tests at Z peak, is it still possible
to find or to constrain NP effects from the
fermionic channels $e^+e^- \to f\bar f$ at higher energies?
The answer is "Yes" because NP effects should grow when the energy
increases towards the NP scale, whereas SM contributions should
decrease (because of unitarity constraints).\par
We have established a
method for treating this situation that we called in \cite{Zsr}\\
{\bf The $Z$-peak Subtracted
Representation}.
The main idea is that
of expressing the various effects in the form of a once-subtracted
dispersion integral, and of fixing the necessary subtraction constants
by suitable model-independent LEP 1 results. In this way, we are led
to a compact "representation"
of all observables (cross sections,
FB asymmetries, polarization asymmetries,....) which presents two
main advantages. The first one is to express the New
Physics contributions through convergent integrals. The second one is
that LEP 1 constraints ($M_Z$ and all $Z$ partial widths), as well as
$\alpha(0)$, are automatically incorporated in the
expressions of the observables.\par
 For example, the cross section
for muon production $ \sigma_\mu(q^2)$, at cm energy $\sqrt{q^2}$,
at one loop level takes the form\par

\bqa
 \sigma_{\mu}^{(1)} (q^{2})   \simeq
\left ( \frac{4}{3} \pi \, q^{2} \right ) \,
\biggm\{ \left [ \frac{\alpha}{q^{2}} \right ]^{2} \,
\left [ 1 + 2 {\bf \tilde{\Delta} \alpha (q^{2})}\ \right ] \nonumber\\
 +  \frac{1}{[(q^{2}-M^{2}_{z})^{2} + M^{2}_{z} \Gamma^{2}_{z}]} \,
\left [ \frac{3\Gamma_{\ell}}{M_{z}} \right]^{2} \,
\biggm [ 1 - 2{\bf R(q^{2})}
 -  \frac{16 (1 - 4 s^{2}_{1}) c_{1} s_{1}}
{(1+\tilde{v}_{\ell}^{2} (M_{z}^{2}))} {\bf V(q^{2})}
\biggm ]\ \biggm\}
\eqa

where in the three combinations
\bqa
{\bf \tilde{\Delta} \alpha (q^{2})}
\equiv {\cal R}e (\tilde{F}_{\gamma}(0)
- \tilde{F}_{\gamma} (q^{2}))
\eqa
\bqa
{\bf R(q^{2})} = \tilde{I}_{z}(q^{2}) - \tilde{I}_{z}(M^{2}_{z})
\eqa
\bqa
{\bf V(q^{2})} = {\cal R}e
\left [\tilde{F}_{\gamma z} (q^{2}) - \tilde{F}_{\gamma z} (M^{2}_{z})
\right ]
\eqa
\noindent
respectively associated to photon exchange, Z exchange
and their interference,
a ``subtraction" at the $Z$ peak has been performed. Other $e^+e^-\to
f\bar f$ processes are treated in a similar way and
involve the same types of combinations with an
index $(ef)$, see \cite{RVtraj}.\par

 The departures from 1-loop SM predictions appear in these
three Z-subtracted combinations.
We have considered several
classes of models (Technicolour type, Anomalous gauge boson couplings,
Higher vector boson Z'), computed their contributions to these
functions and obtained the expressions of the various observables
at LEP2
and NLC. Depending on the number of basic parameters of
the models, typical relations among these various functions have been
established. Three examples are given below, for the Technicolour-type
of models \cite{RVstrong}
\bqa  V^{TC}(q^2) \equiv ({1-2s^2_1\over2s_1c_1})({q^2-M^2_Z\over
q^2})({M^2_V\over M^2_V-M^2_Z})\tilde{\Delta} \alpha^{TC} (q^{2}) \ \ ,
\eqa
for anomalous gauge boson couplings \cite{Zsr}
\bqa \tilde{\Delta} \alpha^{AGC} (q^{2}) \equiv -({q^2\over q^2-M^2_Z})
[ R^{AGC}(q^{2})+{2s^2_1-1\over c_1s_1}V^{AGC}(q^2)] \ \ , \eqa
and for general $Z'$ exchange \cite{RVtraj}
\bqa  [V^{Z'}(q^2)]^2\equiv -({q^2-M^2_Z\over q^2}) R^{Z'}(q^{2})
\tilde{\Delta} \alpha^{Z'} (q^{2}) \ \ .
\eqa
 We have then translated them in the form of trajectories
in the space of the observables. As one can see from
Fig.1a-b this should allow a clear disentangling of these classes
of models. In Fig.1c,d we have illustrated how various types of $Z'$
models distribute inside the 2-dimensional space of $\sigma(e^+e^-\to
\mu^+\mu^-)$ and $A^{\mu}_{FB}$.\par
In \cite{RVtraj} trajectories in the space of other observables
(including hadronic ones) have also been obtained.

\section{$e^+e^-\to W^+W^-$ at LEP2 and NLC}

This process is known since a long time to be the first
place for testing the
gauge couplings $\gamma WW$ and $ZWW$\cite{Gaemers}. After several
other studies in the past, a model independent analysis has been recently
made\cite{Bilenky}. It involves the seven possible forms for anomalous
3-boson couplings, consistent with Lorentz and $U(1)_{EM}$ invariances.
A methodology has been proposed\cite{BMT}, \cite{Bilenky},
\cite{IJMP}, \cite{GRS} for
disentangling these various forms by using the W angular distribution
and the W spin density matrices measurable through the decay
distributions into fermion pairs.
Discovery limits for individual couplings and contour plots for the
multiparameter cases have been obtained.\par
If we now stick to the assumptions about NP explained in the
introduction, and restrict to CP-conserving $dim=6$ operators, only
three independent operators $\O_W$, $\O_{B\Phi}$ and $\O_{W\Phi}$
appear. They contribute to five of the
$\gamma WW$ and $ZWW$ anomalous couplings, namely $\delta_Z$,
$x_{\gamma}$, $x_Z$, $\lambda_{\gamma}$ and $\lambda_Z$,
satisfying the
relations \cite{g.i.}, \cite{Grosse}

\bq  x_Z = -{s\over c}x_{\gamma}  \eq

\bq   \lambda_{\gamma} = \lambda_Z \eq

At LEP2 the
sensitivity to these couplings is expected to be at the level of
$10^{-1}$.
At NLC the sensitivity is expected to increase according to
the
scaling law $\sqrt{s.L}$ where $s$ is the square of the $e^-e^-$ energy
and $L$ the luminosity of the machine. For example at 1 TeV with
$80fb^{-1}$ one should reach a sensitivity of the order of
$10^{-3}$.\par
Using the unitarity relations \cite{unit} this means that NP scales
of about 1.5 TeV should be reached for these operators at LEP2. At
a 1TeV NLC they reach about 10 TeV.

\section{Production $HZ$ and $H\gamma$ at LEP2 and NLC}
If the Higgs mass is low enough, the first possibility of doing tests
of Higgs boson properties will be offered by LEP2 (and later on by
NLC) with $e^+e^-\to HZ$
and $e^+e^-\to H\gamma$.  The SM allows for $e^+e^-\to Z\to HZ$ at
tree level and for $e^+e^-\to \gamma, Z\to H\gamma$ at 1-loop level.
We have concentrated\cite{hz} our analysis to NP effects due to the five
operators  $\O_{UB}$,
$\O_{UW}$, $\ol{\O}_{UB}$, $\ol{\O}_{UW}$ and $\O_{\Phi2}$ which create
anomalous $HZZ$, $HZ\gamma$ and $H\gamma\gamma$
couplings.
The effects of the
other operators should be already largely constrained through the
$e^+e^-\to W^+W^-$ channel. The search of NP should proceed by doing
precision measurements of $HZ$ production and by looking for signals of
$H\gamma$ production, the SM rate for this second process being
apparently too small to be observable.\par
The precision tests of $e^+e^-\to HZ$ that we propose
\cite{hz} should consist in
measuring accurately the angular distribution $d\sigma/dcos\theta$ and
the $Z$ spin density matrix elements through the dependence in the
azimuthal angle $\phi_f$ between the $Z$ production plane and
the $Z\to f\bar f$ decay plane
($f$ being a charged lepton or a $b$ quark).
Four azimuthal asymmetries $A_{14}$, $A_{12}$, $A_{13}$ and $A_8$
respectively associated to the $cos\phi_f$,
$sin\phi_f$, $sin2\phi_f$ and $cos2\phi_f$dependences
are especially suitable for
this search. With these five observables it is in principle possible to
disentangle the effects of the five NP operators.\par
At LEP2, the number of events would be too small (about 200 raw ones) to
allow for such a detailed analysis. Only $d\sigma/dcos\theta$ will be
measurable (see Fig.2a,b)
and will give a meaningful constraint on one combination of
NP couplings namely $d_{ZZ}=dc^2_W+d_B s^2_W$ and $|f_{\phi 2}|$.
Assuming $m_H=80GeV$ at
$192GeV$ the sensitivity limit is $d=0.015$ or $d_B=0.04$ which
corresponds through unitarity relations to NP scales of 14 and 5 TeV
respectively. In the case of  $\O_{\Phi2}$ the observability
limit is $|f_{\phi 2}| \simeq 0.01$, which means
$\Lambda_{NP}\simeq 6-7 TeV$.
This is not too bad for a first exploration of the Higgs
sector!\par

At NLC, for example at $1TeV$
with a few thousands of events, see Fig.2c,
the sensitivity  now reaches  $|f_{\phi 2}|\simeq 0.004$, $|d|\simeq
0.005$, $|d_B|\simeq 0.015$ corresponding to NP scales of 10 , 40
, 8 TeV respectively. The disentangling of the five
couplings is now conceivable through the study of the
azimuthal asymmetries, which depend on four other combinations of
$|f_{\phi 2}|$,
$d_{\gamma Z}=s_W c_W (d-d_B)$
and the CP-violating ones $\bar
d_{ZZ}$ and $\bar d_{\gamma Z}$.
This should be possible down
to the percent level, see Fig.2d,e,f and \cite{hz}.\par
A preliminary analysis of $H\gamma$ production is in progress
\cite{hz}. The
angular distribution $d\sigma/dcos\theta$ of this process is sensitive
to a new combination of couplings $d_{\gamma\gamma}=ds^2_W+d_B c^2_W$.
So it can be used to disentangle $d$ from $d_B$ without going through
the difficult analysis of the $Z$ spin density matrix. Only the search
for CP-violating effects would then motivate it.

\section{Tests in $\gamma\gamma$ collisions}

 New possibilities  for testing the bosonic sector will be offered at
high energy $e^+e^-$ colliders. Through the laser backscattering method
\cite{laser} intense and high energetic photon beams will be available.
For our purpose two types of processes will be interesting, boson pair
production and single Higgs production in $\gamma\gamma$ collisions.\\

{\bf a) Boson pair production in $\gamma\gamma$ collisions}

We have considered\cite{ggvv} the following five processes
$\gamma \gamma \to W^+W^-$,  $\gamma \gamma \to ZZ$,
$\gamma \gamma \to \gamma Z$,
$\gamma \gamma \to \gamma \gamma$, and also $\gamma \gamma \to HH$.
They are sensitive to the seven operators
$\O_W$, $\O_{W\Phi}$, $\O_{B\Phi}$, $\O_{UW}$, $\O_{UB}$,
$\ol{\O}_{UW}$, $\ol{\O}_{UB}$.
The first three of them
induce anomalous triple gauge boson couplings, while
the remaining four create  anomalous CP conserving and CP violating
Higgs couplings.\par
We have shown\cite{ggvv}
that the $p_T$ distribution of one of the final boson
$B_3$ and $B_4$ provides a convenient way of looking for NP effects.
 The process $\gamma\gamma \to W^+W^-$ receives a tree level
contribution from SM, and NP contribution from the seven operators. The
processes $\gamma\gamma \to ZZ$ and $\gamma\gamma \to HH$
receive no tree level SM contribution
but NP contributions from six operators ($\O_W$ is excluded). The two
processes $\gamma\gamma \to \gamma\gamma$ and $\gamma Z$ also receive
no tree level contribution from SM but NP contributions from only the
four operators $\O_{UW}$, $\O_{UB}$,
$\ol{\O}_{UW}$, $\ol{\O}_{UB}$, typical of the scalar sector.\par
We have studied the sensitivity to each of these operators in the
various channels and the ways to disentangle their effects. See
Fig.3a-d.\par
Table 3 summarizes the observability limits (lower bounds for
the couplings leading to observable effects) expected for each
operator by assuming that for $W^+W^-$ production a departure
of 5\% as compared to the SM prediction in the high $p_T$ range,
will be observable. Combining these
bounds with the unitarity relations \cite {unit} we
obtain the upper bounds on the related NP scale which are
indicated in parentheses in Table 3. The luminosities assumed for a
0.5, 1, 2 TeV collider are 20, 80, 320 $fb^{-1}$ respectively.

\begin{center}
\begin{tabular}{|c|c|c|c|c|c|c|c|c|} \hline
\multicolumn{9}{|c|}{Table 3: Observability limits based on
the $W^+W^-$ channel}  \\
\multicolumn{9}{|c|}{to the anomalous couplings and the related
NP scales $\Lambda_{NP}$
(TeV).} \\ \hline
\multicolumn{1}{|c|}{\null} &
\multicolumn{2}{|c|}{\null} &
\multicolumn{2}{|c|}{\null} &
\multicolumn{2}{|c|}{\null} &
\multicolumn{2}{|c|}{\null}  \\
\multicolumn{1}{|c|}
{ \null} &
 \multicolumn{2}{|c|}{ $\O_W$} &
    \multicolumn{2}{|c|}{ $\O_{UW}$ or $\ol{\O}_{UW}$ } &
       \multicolumn{2}{|c|}{ $\O_{UB}$ or $\ol{\O}_{UB}$ } &
    \multicolumn{2}{|c|}{ $\O_{B\Phi}$ or $\O_{W\Phi} $ } \\ \hline
\null & \null & \null & \null & \null &\null &\null &\null &
\null \\
\multicolumn{1}{|c|}{$2E_e$(TeV)} &
  \multicolumn{1}{|c|}{$|\lambda_W|$} &
   \multicolumn{1}{|c|}{$\Lambda_{NP}$} &
   \multicolumn{1}{|c|}{$|d|$ or ${|\ol{d}|}$} &
   \multicolumn{1}{|c|}{$\Lambda_{NP}$} &
    \multicolumn{1}{|c|}{$|d_B|$ or  $|\ol{d}_B|$ } &
    \multicolumn{1}{|c|}{$\Lambda_{NP}$} &
      \multicolumn{1}{|c|}{$|f_B|$ or $|f_W|$} &
      \multicolumn{1}{|c|}{$\Lambda_{NP}$}     \\ \hline
 0.5 & 0.04 & 1.7 & 0.1 &2.4  & 0.04 &4.9 & 0.2 & 1.8, 1 \\
 1  & 0.01 &3.5 & 0.04 &5.5 & 0.015 & 9  & 0.05 & 3.5, 2 \\
 2 & 0.003 & 6.4 & 0.015 & 17 & 0.005 & 17 & 0.015 &6.5, 3.6  \\ \hline
\end{tabular}
\end{center}
\noindent

At 2 TeV, in the case of the operators  $\O_{UB}$, $\O_{UW}$,
$\ol{\O}_{UB}$, $\ol{\O}_{UW}$ slightly better limits could in
principle be obtained by using the $\gamma\gamma \to ZZ$ process.
Demanding for example that the NP contribution to this process
reaches the level of the SM result for $\gamma\gamma \to ZZ$
\cite{ggZZloop},  we can decrease the
limiting value for the couplings $|d|$
(or $|\ol{d}|$) and $|d_B|$ (or $|\ol{d}_B|$)
down to 0.01 and 0.003 respectively. This means that a 2 TeV
collider is sensitive to NP scales of about 20 TeV.
A more precise analysis using  realistic uncertainties for the
detection of the ZZ channel and taking into
account the interference between the SM and the NP
contributions,  could probably improve these limits.
This is left for a future work.\par

We also found\cite{ggvv}
that complete disentangling is possible by analyzing final spin states,
i.e. separating $W_T(Z_T)$ from $W_L(Z_L)$ states.
Identification of CP violating terms requires full $W$ or $Z$
spin density matrix reconstruction from their decay
distributions\cite{GRS, hz} or
analyses with linearly polarized photon beams\cite{Kraemer}.\\
\vspace{0.3cm}

{\bf b) Single Higgs production in $\gamma\gamma$ collisions}

We now show\cite{ggh} that single Higgs production in $\gamma\gamma$
collisions through laser backscattering should provide the best way to
look for New Physics (NP) effects inducing anomalous Higgs couplings.
The Standard contribution to $\gamma\gamma\to H$ only occurs at
1-loop. With the high luminosities expected
at linear $e^+e^-$ colliders, a large number of Higgs bosons
should be produced. The
sensitivity to anomalous $H\gamma\gamma$
couplings is therefore very strong. NP contributions to this coupling
only arises through the four operators $\O_{UB}$,
$\O_{UW}$, $\ol{\O}_{UB}$ and $\ol{\O}_{UW}$.\par
We have first studied\cite{ggh}
the sensitivity of the production cross section
$\sigma(\gamma\gamma\to H)$ for the typical NLC
energies.
With the aforementioned designed luminosities,
one gets a few thousands
of Higgs bosons produced in the light or intermediate mass range, see
Fig.4a,b for the case of a 1 TeV NLC.
Assuming conservatively an
experimental detection accuracy of about 10\% on the
production rate, one still
gets an observability limit of the order of $10^{-3}$,
$4.10^{-3}$, $3.10^{-4}$, $10^{-3}$
for $d$, $\bar d$,
$d_B$ and $\bar d_B$ respectively. The corresponding constraints
on the NP scale derived on the basis of the unitarity relations
\cite{unit},
are then very high, i.e. 200, 60, 60 and 30 TeV respectively.\par
The disentangling of the four operators $\O_{UB}$,
$\O_{UW}$, $\ol{\O}_{UB}$ and $\ol{\O}_{UW}$
is possible by considering the Higgs decay branching
ratios. Concerning this, we remark that $H\to b\bar
b$ is not affected by the aforementioned seven purely bosonic
operators describing NP. Neither $H\to
WW,\, ZZ$ are particularly sensitive to such an NP,
since it is masked by
strong tree level SM contributions. On the other hand, the processes
$H \to \gamma\gamma$ and  $H \to \gamma Z$ who receive tree level
contributions from the above anomalous couplings and only one
loop ones from SM, are most sensitive to NP.
Thus their ratio to the dominant Higgs decay
mode, which depending on the Higgs mass may either be the $b\bar b$
or the $WW,\, ZZ$ modes, provide a very sensitive way to further help
disentangling among the CP conserving operators $\O_{UB}$ and
$\O_{UW}$. This way, using $H \to \gamma \gamma$ or $H \to
\gamma Z$,  couplings even weaker than $10^{-3}$ could
be observable. For comparison we note that the corresponding
sensitivity limit from $H\to WW,\, ZZ$ is at the 10\% level.
 \par
Most spectacular for the disentangling of the various operators
seem to be the ratios $WW/ZZ$ and $\gamma\gamma/\gamma Z$, shown in
Fig.5a,b.
The first one, which is applicable in the intermediate
and high Higgs mass range, allows to disentangle $\O_{UB}$
from $\O_{UW}$ down to values of the order of $10^{-1}$, whereas the
second one, applicable in the light Higgs case, is sensitive to
couplings down to the $10^{-3}$ level or less.\par
Note that these properties are independent of the production modes and
can be used for disentangling anomalous Higgs couplings in any other
process.\par
The identification of CP-violating terms is not directly
possible except for the remark that, contrarily to the
CP-conserving terms, there can be no intereference with the tree
level SM contributions. Thus a CP violating
interaction cannot lower the value of the widths through a
destructive intereference. A direct identification of CP
violation requires
either an analysis of the W or Z spin density matrix through
their fermionic decay distributions \cite{GRS, hz}, or
the observation of a suitable asymmetry with linearly polarized
photon beams \cite{Kraemer}.

\section{Concluding words}

High precision measurements at Z peak on the one hand,
technical problems and deficiencies of SM as well as
dynamical models for NP residual effects on the other hand,
suggest a hierarchy among
the sectors where NP manifestations should first appear. This hierarchy
favors the scalar sector and at a lower level the gauge sector and the
heavy quark sector. The light fermionic sector should be disfavored.\par
According to this hierarchy
and to several dynamical and symmetry properties
we have classified the operators
used to construct the effective lagrangian representing the NP effects
at energies much below the NP scale. We have established
unambiguous relations between the coupling constants and the NP scale
through unitarity constraints.
We have shown how higher Z' vector
bosons effects can be
integrated into modified photon and Z exchange amplitudes.\par
We have then
examined the search for such effects at present and future colliders.
We have first looked at virtual effects of NP in $e^+e^-\to f\bar f$.
We have analyzed LEP1/SLC high precision tests
in the various sectors in
order to give model independent constraints on Z' couplings and other
virtual NP effects due to, for example, residual bosonic
and residual heavy quarks
interactions described by effective lagrangians. We have established
a method for representing
$e^+e^-\to f\bar f$ at high $q^2$, taking automatically
into account Z peak results in
all observables. We have applied it to LEP2 and NLC energy ranges for
several types of virtual NP effects (Technicolour type of resonances,
anomalous gauge couplings, general type of $Z'$)
and we have shown that each model is
caracterized by a trajectory in the space spanned by the
departures from SM predictions to the usual observables. So a clear
disentangling of the models should be possible.\par
We have then concentrated on tree level effects
of the effective lagrangian
in the pure bosonic sectors and their tests at LEP2 and NLC in $e^+e^-$
and $\gamma\gamma$ collisions. We have shortly discussed
previous works on the
search for anomalous 3-boson couplings in $e^+e^-\to W^+W^-$. We have
presented new results on $e^+e^- \to HZ, H\gamma$ and the tests that
these processes provide for
anomalous $HZZ$, $HZ\gamma$ and $H\gamma\gamma$ couplings. If the Higgs
is light enough, very sensible results could already be obtained at
LEP2 and then, at NLC, a complete disentangling should be possible.
Through laser backscattering method, the processes $\gamma\gamma\to
W^+W^-, ZZ, Z\gamma, \gamma\gamma$ and $HH$ allow to test
3-boson, 4-boson and Higgs-gauge boson couplings. The most powerful
tests of $H\gamma\gamma$ will however be available through single Higgs
production in $\gamma\gamma\to H$. Independently of the Higgs production
process, ratios of Higgs decay widths can be used to disentangle the
operators involved. The ratio $H\to\gamma\gamma / H\to\gamma Z$ has been
shown to be especially sensitive even for very low couplings.\par
The landscape for these future tests is the following. At present, from
LEP1/SLC the value of the NP scale which can be reached through virtual
effects of blind operators is of the order of 0.35 TeV. The process
$e^+e^-\to W^+W^-$ will allow a direct access to 3-gauge boson couplings
up to an NP scale of about 1.5 TeV at LEP2 and 10 TeV at NLC. The
processes $e^+e^-\to HZ, H\gamma$ will test the Higgs-gauge boson
couplings up to 5-14 TeV at LEP2 and 10-50 TeV at NLC. Finally at NLC
with the laser induced $\gamma\gamma$ collisions one should reach
scales in the 10 TeV range with $\gamma\gamma\to VV$ but in the 100 TeV
range with the single Higgs production process $\gamma\gamma\to H$.
This range of scales covers a large domain of theoretical models
of NP. This can let us hope that we will learn a lot about the origin of
gauge symmetry breaking and the mass generation mechanism.

\vspace{0.5cm}
{\bf Acknowledgements}: This report is based on several works
done in collaboration with D. Comelli, G.J. Gounaris, J. Layssac,
J.E. Paschalis, G. Tsirigoti, N.D. Vlachos and C. Verzegnassi.
We especially thank Jacques Layssac for his help in the preparation of
the figures.

\newpage

\centerline { {\bf Figure Captions }}

{\bf Fig.1} Trajectories in the space of relative departures from SM for
leptonic observables in $e^+e^-\to\mu^+\mu^-$ for various types of
models.\\
(a) 3-dimensional trajectories for general $Z'$\\
(b)Technicolour models (TC) and anomalous gauge couplings (AGC) \\
(c) 2-dimensional trajectories for specific $Z'$ models,
$E_6(-1<cos\beta<+1)$, $RL(\sqrt{2\over3}<\alpha_{RL}<\sqrt2)$ \\
(d) compositeness inspired $Y$, $Y_L$, $Z*$ models.

{\bf Fig.2} $HZ$ production at LEP2 and NLC for $m_H=80 GeV$.\\
(a) Cross section for $e^+e^-\to HZ$ versus $e^+e^-$ total
energy for  SM and NP contributions due to  $\O_{UB}$,
$\O_{UW}$, and $\O_{\Phi2}$.\\
(b) Angular distribution of $e^+e^-\to HZ$
versus $|cos\theta|$ at 192 GeV,  SM and NP contributions
due to  $\O_{UB}$,
$\O_{UW}$, and $\O_{\Phi2}$. The number of
events obtained with an integrated luminosity of 300 $pb^{-1}$
is also indicated.\\
(c) Angular distribution of $e^+e^-\to HZ$
versus $|cos\theta|$ at NLC(1
TeV), SM and NP
contributions due to  $\O_{UB}$,
$\O_{UW}$, and $\O_{\Phi2}$.  The number of
events obtained with an integrated luminosity of 80 $fb^{-1}$
is also indicated. \\
(d) Azimuthal asymmetry $A_{14}$ for  $Z\to b\bar b$,
versus total $e^+e^-$ energy, SM and NP contributions due to
$\O_{UW}$.
\\
(e) Azimuthal asymmetry  $A_{12}$ for  $Z\to b\bar b$,
versus total $e^+e^-$ energy,
SM and NP contributions due to $\ol{\O}_{UW}$.
\\
(f) Azimuthal asymmetry  $A_{13}$ versus total $e^+e^-$ energy,
SM and NP contributions due to $\ol{\O}_{UW}$.

{\bf Fig.3} Sensitivity of the transverse momentum ($p_T$)
distribution $d\sigma/d p_T$
at 1 TeV for various operators.\\
(a) $\O_W$ in $\gamma\gamma \to
W^+W^-$.\\
(b) $\O_{B\Phi}$ and $\O_{W\Phi}$
in $\gamma\gamma \to W^+W^-$, $ZZ$, $HH$.\\
(c) $\O_{UW}$ and $\ol{\O}_{UW}$ in
$\gamma\gamma \to W^+W^-$, $ZZ$, $\gamma Z$, $\gamma\gamma$, $HH$.\\
(d) $\O_{UB}$ and $\ol{\O}_{UB}$
in $\gamma\gamma \to W^+W^-$, $ZZ$, $\gamma Z$, $\gamma\gamma$, $HH$.

{\bf Fig.4} Cross sections for Higgs production in
$\gamma\gamma$ collisions
from laser backscattering at a 1 TeV
  $e^+e^-$ linear collider. The expected number of events per year
for an integrated luminosity of $80 fb^{-1}$,
is also
indicated.\\
(a) Standard prediction
(solid line),
 with $d =+ 0.01$ (long dashed),  $d =- 0.01$ (dashed - circles),
  $d =+ 0.005$ (short dashed),  $d =- 0.005$ (dashed),
  $d =+ 0.001$ (dashed - stars), and $d =- 0.001$ (dashed - boxes).\\
(b) Standard prediction
(solid line),
 with $\bar d = 0.01$ (long dashed),
  $\bar d = 0.001$ (short dashed),
  $\bar d = 0.005$ (dashed).

{\bf Fig.5}  Ratios of Higgs decay widths for $m_H=0.2 TeV$
 versus coupling constant values,\\
(a) $\Gamma(H\to WW)/\Gamma(H\to ZZ)$ ,\\
(b) $\Gamma(H\to\gamma \gamma)/\Gamma(H\to \gamma Z)$ ,\\
with $d>0 $ (solid), $d<0 $ (short dashed), $d_B>0$
(dashed), $d_B<0$ (long dashed),
 $\bar d $ (dashed-circles), $\bar d_B$ (dashed-stars).


\newpage

\begin{thebibliography}{99}


\bibitem{LEP1} D. Schaile, presented at the
27th Int. Conf. on High Energy Phys., Glasgow (1994). J. Erler and
P. Langacker, Univ of Pennsylvania preprint UPR-0632T (1994).
%
\bibitem{Haber} J. Gunion, H. Haber, G. Kane and S. Dawson, The
Higgs Hunter's guide, Addison-Wesley, Reading 1990.
%
\bibitem{Buchmuller} W. Buchm\"{u}ller and D. Wyler,
\np{B268}{1986}{621}; C.J.C. Burgess and H.J. Schnitzer,
\np{B228}{1983}{454}; C.N. Leung, S.T. Love and S. Rao
\zp{C31}{1986}{433}.
%
\bibitem{unit} G.J. Gounaris, J. Layssac and F.M. Renard,
\pl{B332}{1994}{146}.
%
\bibitem{uni2} G.J. Gounaris, J. Layssac, J.E. Paschalis
and F.M. Renard, Montpellier preprint PM/94-28, Z. Phys. C66(1995)619.
%
%
\bibitem{model}G.J. Gounaris, F.M. Renard and G. Tsirigoti,
Phys. Lett. B350(1995)212.
\bibitem{DeR} A. De R\'{u}jula, M.B. Gavela, P. Hernandez
and E. Masso,\np{B384}{1992}{3}.
%
\bibitem{g.i.} G.J. Gounaris and F.M. Renard, \zp{C59}{1993}{133}.

\bibitem{SU2V}J.L. Kneur and  D. Schildknecht,\np {B357} {357}
{1991};
J.L. Kneur, M. Kuroda and D. Schildknecht,\pl {B262} {93} {1991}.
M. Bilenky \etal\ , \pl{B316}{1993}{345}

\bibitem{dyn}G.J. Gounaris, F.M. Renard and G. Tsirigoti,
\pl{B338}{1994}{51}.

\bibitem{ZZp} J. Layssac, F.M. Renard and C. Verzegnassi,
\pl{B287}{1992}{267},\zp{C53}{1992}{97}.

\bibitem{ZZplep2} J. Layssac, F.M. Renard and C. Verzegnassi,
contribution to the LEP2 workshop (1995).

\bibitem{RVtraj} F.M. Renard and C. Verzegnassi, preprint PM/95-35
(1995).

\bibitem{Zpfnal}CDF Collaboration, Abe et al., FERMILAB-PUB-94-198-E
(1994).

%
\bibitem{Hag}
K. Hagiwara, S. Ishihara, R. Szalapski and D. Zeppenfeld,
\pl{B283}{1992}{353} and \pr{D48}{1993}{2182}.

\bibitem{topbb} G.J. Gounaris, F.M. Renard and C. Verzegnassi,
Preprint PM/94-44 and THES-TP 95/01, hep-ph/9501362n to appear
in Phys. Rev. D.
%
\bibitem{dirtop} G.J. Gounaris et al, in preparation.

\bibitem{AB} G. Altarelli, R. Barbieri, F. Caravaglios, \pl {B314} {357}
{1993}; M. Peskin and T. Takeuchi, \pr {D46} {381} {1992};
B.W. Lynn, M.E. Peskin and R. Stuart in "Physics at
LEP", J. Ellis and R. Peccei eds., CERN 86-02 (1986), Vol 1;
{Hollik}M. Consoli and W. Hollik in ``Physics at LEP I",
G. Altarelli, R. Kleiss, C. Verzegnassi eds. CERN 89-09 (1989).

\bibitem{vertex} J. Layssac, F.M. Renard and C. Verzegnassi,
\pr{D49}{1994}{3650}.

\bibitem{D} F.M.Renard and C.Verzegnassi, \pl {B260}{1991}{225}


\bibitem{LEPpol} A.Blondel, B.W.Lynn, F.M. Renard and C. Verzegnassi,
 \np {B304} {438} {1988}.
%
\bibitem{Burgess} C.P. Burgess and D. London, \pr{D48}{1993}{4337}.


\bibitem{Comelli} D. Comelli, F.M. Renard and C. Verzegnassi, \pr
{D50}{1994}{3076}.

\bibitem{RV} F.M. Renard and C. Verzegnassi, \pl{B345}{1995}{500}.

\bibitem{Zsr} F.M. Renard and C. Verzegnassi, CERN-TH.7485/94 to appear
in Phys.Rev.


\bibitem{RVstrong} J. Layssac, F.M. Renard and C. Verzegnassi,
\pr{D49}{1994}{2143}.

%
\bibitem{Gaemers}K.J.F. Gaemers and G.J. Gounaris,
\zp{C1}{1979}{259}; K. Hagiwara, R. Peccei, D. Zeppenfeld and K.
Hikasa, \np{B282}{1987}{253}.

%
\bibitem{BMT} G.J.Gounaris et al, in Proc. of the Workshop on
    $e^+e^-$ Collisions a  500 GeV: The Physics Potential,
    DESY 92-123B(1992), p.735, ed. P.Zerwas;
    M.Bilenky et al, {\bf{B419}} (1994) 240.
%
\bibitem{Bilenky}M. Bilenky, J.L. Kneur, F.M. Renard and
D. Schildknecht, \np{B409}{1993}{22} and
{\bf{B419}} (1994) 240.
%

\bibitem{IJMP} G. Gounaris, J. Layssac, G. Moultaka and F.M. Renard,
\ijmp{A8}{1993}{3285}.

%
\bibitem{GRS} G.J. Gounaris, D. Schildknecht and F.M. Renard,
\pl{B263}{1993}{143}.

\bibitem{Grosse} C. Grosse-Knetter, I. Kuss and D. Schildknecht,
\zp{C60}{1993}{375}.

\bibitem{hz} G. Gounaris, F.M. Renard and N.D. Vlachos, preprint
PM/95-30, THES-TP 95/08.

%
\bibitem{laser}  I. Ginzburg et al, Nucl. Instrum. Methods
${\bf 205}$(1983)47; ${\bf 219}$(1984)5.
%
\bibitem{ggvv} G.J. Gounaris, J. Layssac and F.M. Renard, preprint
PM/95-11, THES-TP 95/06.

%
\bibitem{ggZZloop} G.V. Jikia, \np{B405}{1993}{24};
D.A. Dicus and C. Kao, \pr{D49}{1995}{1265};
 M.S. Berger, \pr{D48}{1993}{5121}.

%
\bibitem{Kraemer} M. Kr\"{a}mer, J. K\"{u}hn, M.L. Stong and P.M.
Zerwas, \zp{C64}{1994}{21}.
%
\bibitem{ggh}  G.J. Gounaris and F.M. Renard, preprint
PM/95-20, THES-TP 95/07.









\end{thebibliography}
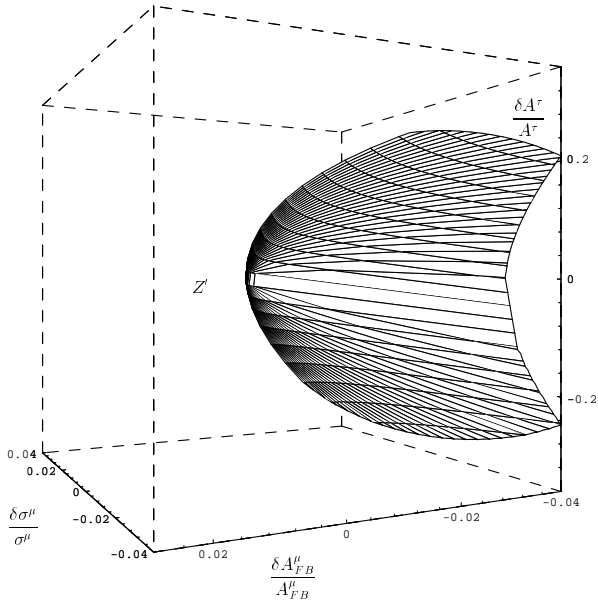
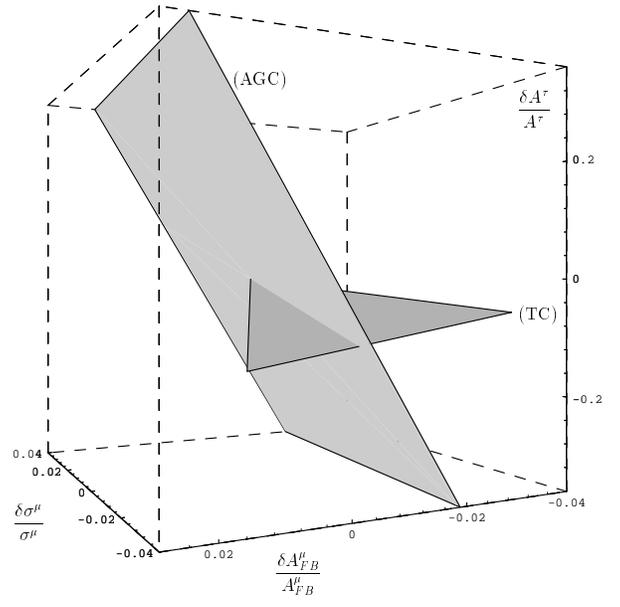
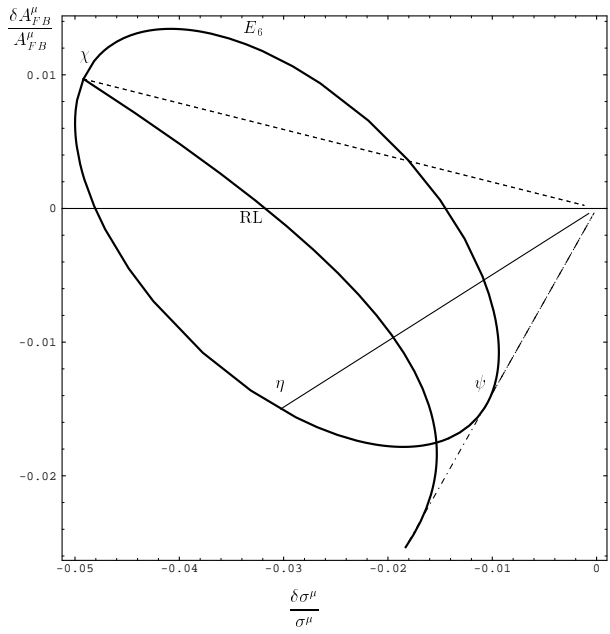
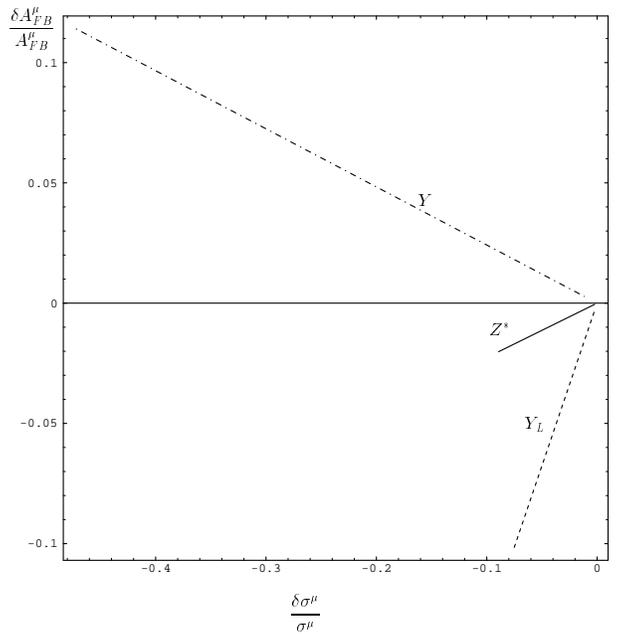
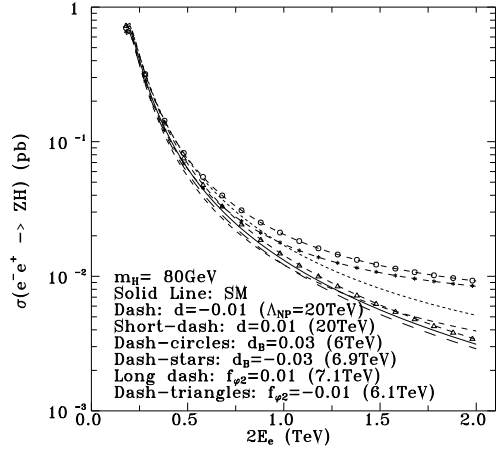
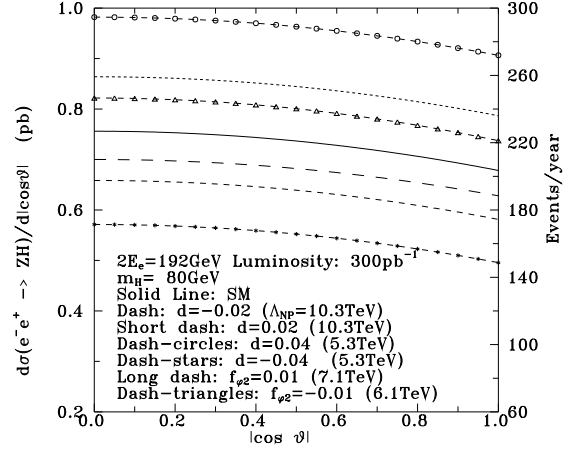
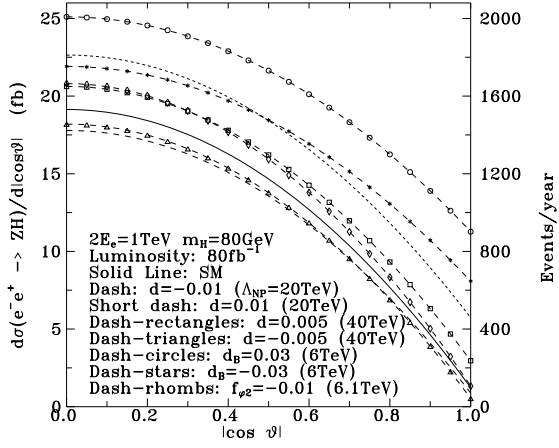
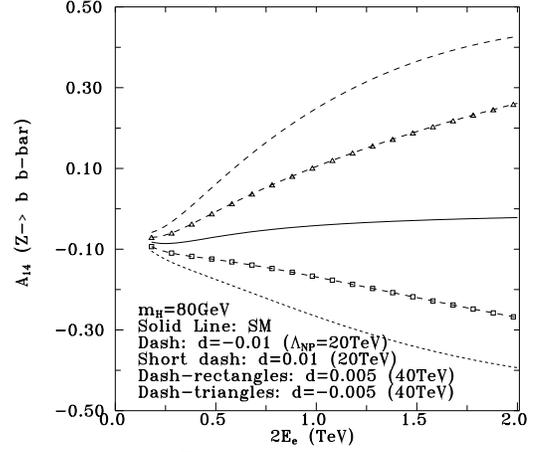
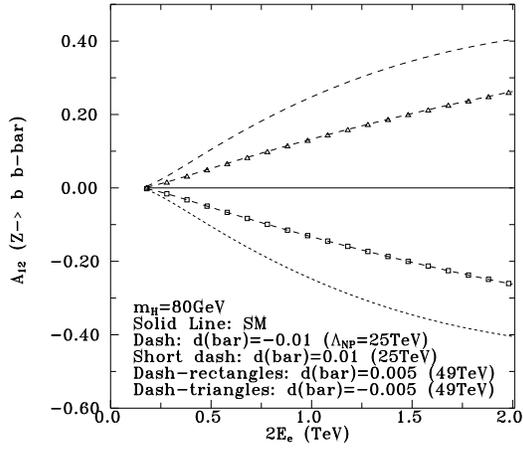
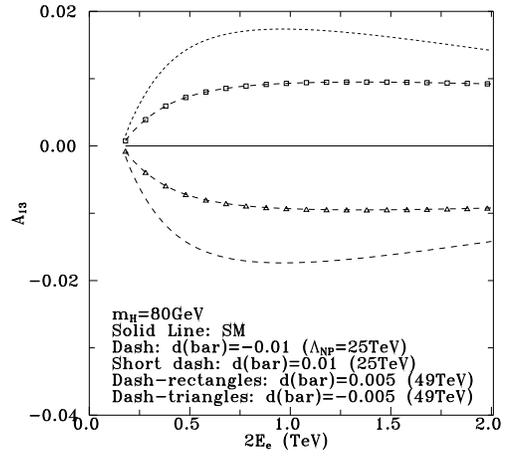
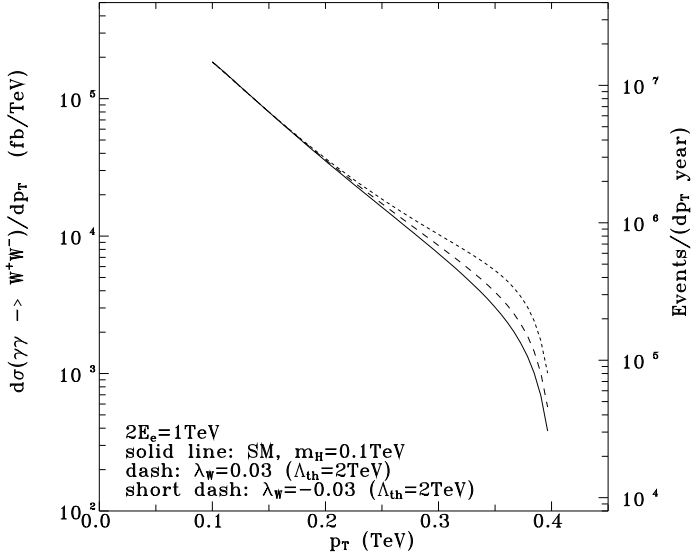
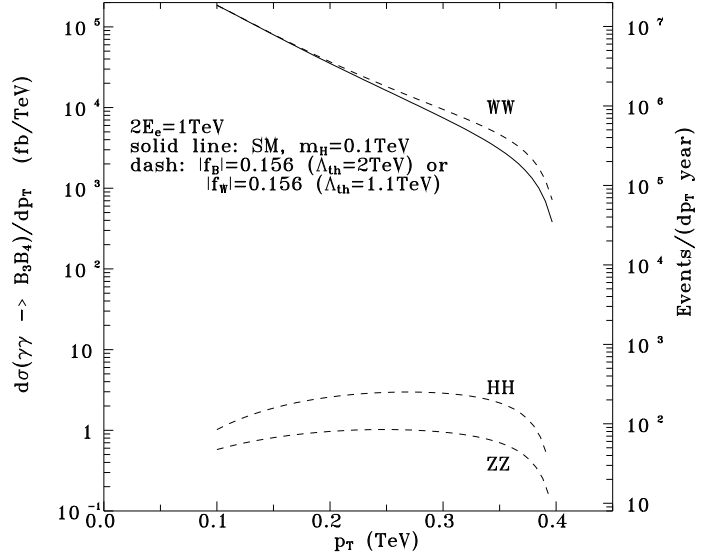
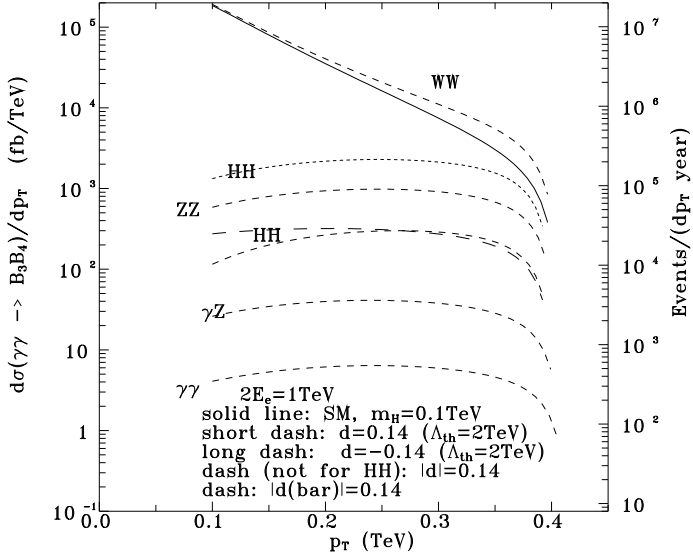
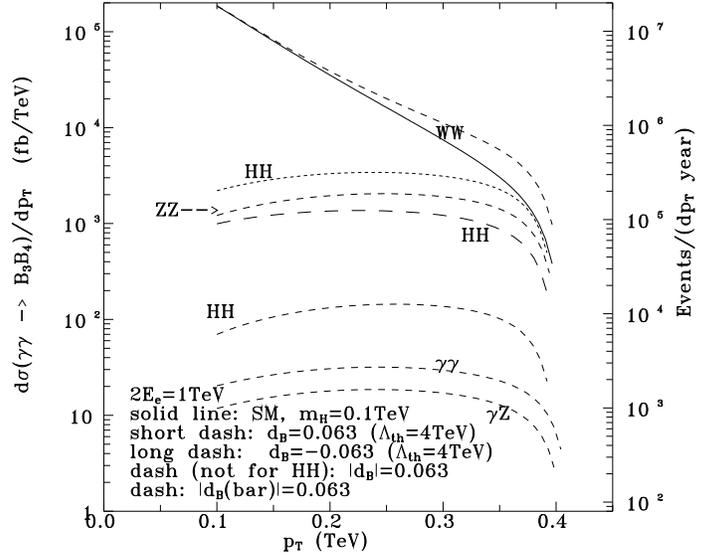
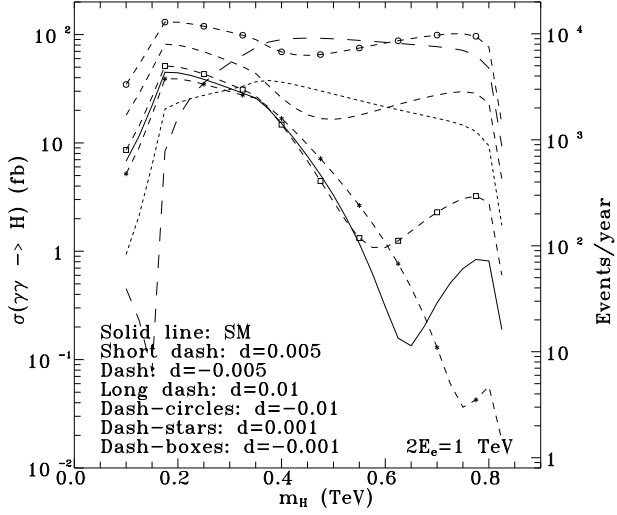
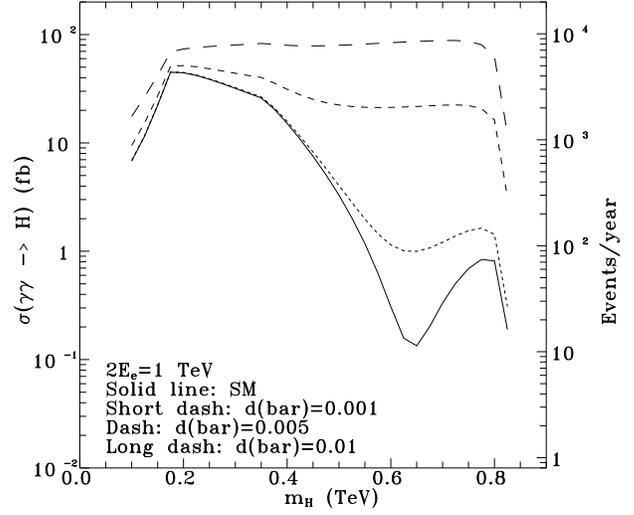
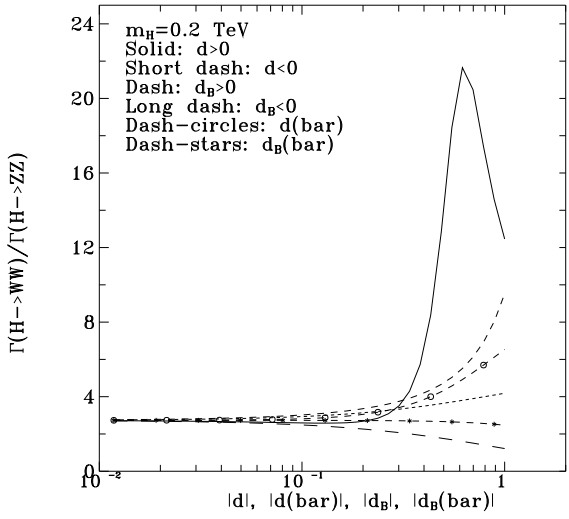
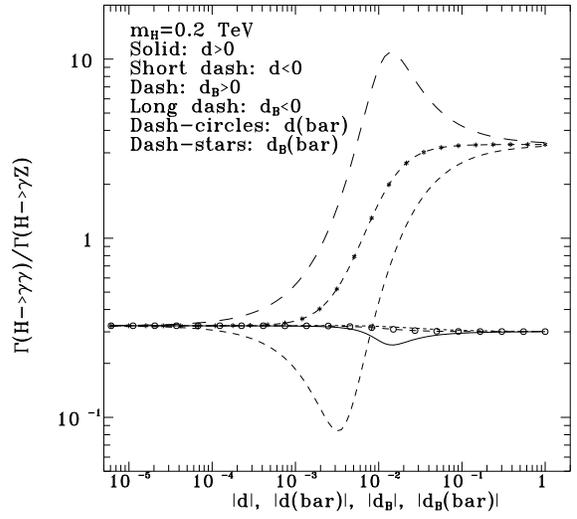
\end{document}